\documentclass[a4paper,12pt,fleqn]{article}%
\usepackage{amsmath, amssymb}
\usepackage{color}

\numberwithin{equation}{section}

\begin{document}

\title{\textsf{Equivariance on Discrete Space \\ and\\ Yang-Mills-Higgs Model}}
\author{\textsf{Hitoshi Ikemori } \thanks{ikemori@vega.aichi-u.ac.jp}\\Faculty of Economics, Aichi University, \\Nagoya, Aichi 453-8777, Japan
\and \textsf{Shinsaku Kitakado }\thanks{kitakado@ccmfs.meijo-u.ac.jp}\\
Professor Emeritus of Nagoya University
\and \textsf{Yoshimitsu Matsui }\thanks{matsui@vega.aichi-u.ac.jp}\\Faculty of Law, Aichi University,\\Nagoya, Aichi 453-8777, Japan
\and \textsf{Hideharu Otsu } \thanks{otsu@vega.aichi-u.ac.jp}\\Faculty of Regional Policy, Aichi University,\\Toyohashi, Aichi 441-8522, Japan
\and \textsf{Toshiro Sato} \thanks{tsato@sist.chukyo-u.ac.jp}\\School of Engineering, Chukyo University, \\Nagoya, Aichi 466-8666, Japan }
\date{}
\maketitle

\begin{abstract}
We introduce the basic equivariant quantity $Q$ in the gauge theory on the
noncommutative descrete $Z_{2}$ space, which plays an important role for the equivariant
dimensional reduction. If the gauge configuration of the ground state on the
extra dimensional space is described by the equivariant $Q$, then the extra
dimensional space is invisible. Especially, using the equivariance principle,
we show that the Yang-Mills theory on $R^{2}\times Z_{2}$ space is equivalent
to the Yang-Mills-Higgs model on $R^{2}$ space. It can be said that this model
is the simplest model of this type.

\end{abstract}

\newpage

\section{Introduction}

Discovery of Higgs boson has brought about a lot of activities concerned with
the origin of this particle. Although various models have been proposed, the
convincing one does not seem to exist. Higgs boson was introduced as a
particular particle that spontaneously violates the gauge symmetry. Thus, it
is important to investigate the origin of this boson. One candidate could be
to consider the extra dimensions to our real space that we are recognizing and
to reduce the extra dimensions by considering the equivariance of symmetries
involved\cite{Popov,Manton:2010wu,Chatzistavrakidis:2010tq,Madore:1994mj,Harland:2009kg}.

Equivariance implies that the symmetry of the real space is related to that of
the internal space, so that the shifted point in the real space can be sent
back to the original point by the symmetry transformation in the internal
space. Thus the extra dimensional space, even when it is there, could have
been unobserved if the symmetry of the extra space is equivariant to the gauge
symmetry of the real space, that we are actually living in. The gauge fields in
the extra dimensional space are observed as the Higgs fields in the real space.
In other words, the Higgs fields can be considered as the gauge fields in the
extra dimensional space.

Consider, for example, the case of $R^{2}\times S^{2}$, where the real space
is $R^{2}$ and the extra space is $S^{2}$\cite{Popov,Manton:2010wu}.
Equivariance on $S^{2}$ implies that pure Yang-Mills (YM) theory
is recognized as Yang-Mills-Higgs (YMH) model on $R^{2}$. Thus, self dual (SD)
equation on $R^{2}\times S^{2}$ is equivalent to
Bogomol'nyi-Prasad-Sommerfield (BPS) equation on $R^{2}$\cite{Bogomolny:1975de}.
 Invisibility of the extra dimensions is
guaranteed by the fact that configuration of the ground state is equivariant.
Such an equivariant gauge configuration can be constructed using the simplest
equivariant quantity $Q\equiv i\hat{x}_{a}\sigma_{a}$, where $\hat{x}_{a}$ is
three dimensional coordinate describing $S^{2}$ and $i\sigma_{a}/2$ is the
gauge symmetry generator. This also describes the configuration of the so
called \textquotedblleft Witten ansatz\textquotedblright\cite{Witten:1976ck}.

Equivalence of pure YM theory and YMH model through the existence of extra
dimensional space, has been discussed also in other models, with a little more
generality, like in the coset space $S/R$ \cite{Popov}
\cite{Manton:2010wu} model or the fuzzy version of it,
$(S/R)_{F}$ \cite{Chatzistavrakidis:2010tq,Madore:1994mj,Harland:2009kg}.

We have seen a similar extra dimensions for the case of noncommutative $Z_2$, using the method of differential forms\cite{Ikemori:2008pb} (see also \cite{Teo:1997cn}) 
and introducing the coordinates of noncommutative $Z_2$ space\cite{Otsu}.
In those papers, we have shown that difference of vortex and instanton can be considered as the difference that  the space
where they exist is $R^2\times Z_2$, and $R^4$, and that they satisfy the same self dual equation in each space. This implies that 
the vortex may be treated analytically as the instanton. 
It would be expected that such a construction brings us many advantages to understand the solution for the vortex equation. 
However, the concepts of equivariance, so far, was not unambiguous in this approach.

In the present paper we would like to show that equivariance is important in
this case also and we have explicitly defined $Q$ in noncommutative $Z_{2}$ space, thus we
can introduce \textquotedblleft Witten ansatz\textquotedblright\ for the
noncommutative $Z_{2}$ space, which leads to equivalence of SD equation on
$R^{2}\times Z_{2}$ to BPS equation on $R^{2}$.

Dimensional reduction through the use of equivariance principle especially for
the case of $S^{2}$ and SU(2) gauge symmetry, we have to have at
least a larger symmetry group that includes SU(2) as a subgroup. On the other
hand, in the model that we are proposing, the extra space is a
discrete $Z_{2}$, and the relevant symmetry for the equivariance
is the discrete part of the gauge symmetry. Thus, there is no need to consider
a larger gauge symmetry i.e. we could remain with the same gauge symmetry, and
this could be the simplest possible model of this kind.

In the next sections, we recapitulate the arguments of the dimensional
reduction for the YM theory on $R^{2}\times S^{2}$ based on the equivariance
principle. And based on this argument, we consider the YM theory on
$R^{2}\times Z_{2}$. The last section is devoted to discussions.

\section{Equivariance on $R^{2}\times S^{2}$ Model}

As stated in the previous section, extra dimensional space orthogonal to the
real space, can be left unobserved when the
symmetry of the extra dimensional space were equivariant to the gauge symmetry
of the real world. Instead, the gauge field on the extra dimensional space
makes its appearance as the Higgs field in our world. For the invisible extra
dimensional space, the gauge invariant ground state has to be an equivariant
configuration. For example, let us consider SU($2N$) gauge symmetric model on
$R^{2}\times S^{2} $. In this case, $S^{2}$ can become equivariant extra
dimensional space, when the gauge configuration on $S^{2}$ is described in
terms of equivariant basic quantity $Q\equiv i\hat x_{a}\sigma_{a}$.

In order to confirm equivariance, we examine whether the following symmetry
equations, which are due to Forgacs \& Manton\cite{Forgacs:1979zs},
\begin{align}
&\epsilon_{ijk}x_{j}\partial_{k}A_{l}+\epsilon_{ilk}A_{k}-[\mathcal{J}%
_{i},A_{l}]=0,\\
&\epsilon_{ijk}x_{j}\partial_{k}B-[\mathcal{J}_{i},B]=0\nonumber
\end{align}
are satisfied or not. Here $A_{i}$ is a vector and $B$ is a
scalar, $\mathcal{J}_{i}$ is a generator of internal space. For example, as
$Q$ is a scalar, we substitute $B=Q$ and we obtain
\begin{equation}
\epsilon_{ijk}x_{j}\partial_{k}Q=\epsilon_{ijk}x_{j}\partial_{k}(i\hat{x}%
_{l}\sigma_{l})=i\epsilon_{ijk}x_{j}\sigma_{k}.
\end{equation}
As $\mathcal{J}_{i}=i\sigma_{i}/2$, we have
\begin{equation}
\lbrack\mathcal{J}_{i},Q]=\left[i{\frac{\sigma_{i}}{2}},i\hat{x}_{l}\sigma
_{l}\right]=i\epsilon_{ijk}\hat{x}_{j}\sigma_{k},
\end{equation}
thus $Q$ satisfies the symmetry equation. In other words, when the ground
state is described in terms of Q , the existence of $S^{2}$ could have been
unobserved. Also, the gauge symmetry SU($2N$) is reduced to the
smaller symmetry.

As $Q^{2}=-1$, the vectors that can be constructed from Q are
$\partial_{a}Q$ and $Q\partial_{a}Q$. Thus, the most general gauge
configuration can be written as
\begin{equation}
\displaystyle{A_{a}={\frac{i}{2}}(\varphi_{1}-1)\partial_{a}Q+{\frac{i}{2}%
}\varphi_{2}Q\partial_{a}Q}.
\end{equation}
Since
\begin{align}
&\partial_{a}Q=i\partial_{a}(\hat{x}_{b}\sigma_{b})={\frac{i}{r}}(\delta
_{ab}-\hat{x}_{a}\hat{x}_{b})\sigma_{b},\\
&Q\partial_{a}Q=-{\frac{1}{r}}(\hat{x}_{c}\sigma_{c})(\delta_{ab}-\hat{x}%
_{a}\hat{x}_{b})\sigma_{b}=-{\frac{i}{r}}\epsilon_{acd}\hat{x}_{d}\sigma_{c},
\end{align}
we can rewrite $A_{a}$ in terms of these as\cite{Eto}
\begin{equation}
A_{a}={\frac{i}{2r}}\left[  (H^{\dagger}-1)\omega_{ab}+(H-1)\omega
_{ab}^{\dagger}\right]  \sigma_{b},\quad\omega_{ab}\equiv i(\delta_{ab}%
-\hat{x}_{a}\hat{x}_{b}+i\epsilon_{abc}\hat{x}_{c}),
\end{equation}
where
\begin{equation}
\varphi_{1}={\frac{H^{\dagger}+H}{2}},\quad\varphi_{2}={\frac{H^{\dagger}%
-H}{2}}i.
\end{equation}
This is the \textquotedblleft Witten ansatz\textquotedblright\cite{Witten:1976ck}.
In terms of stereographically projected coordinates
\begin{equation}
y={\frac{\hat{x}_{1}-i\hat{x}_{2}}{1-x_{3}}}=e^{-i\varphi}\tan{\frac{\theta
}{2}},\quad\bar{y}={\frac{\hat{x}_{1}+i\hat{x}_{2}}{1-x_{3}}}=e^{i\varphi}%
\tan{\frac{\theta}{2}},%
\end{equation}
$Q$ can be rewritten as
\begin{equation}
Q={\frac{i}{{1+y\bar{y}}}}\left(
\begin{array}
[c]{cc}%
-1+y\bar{y} & 2y\\
2\bar{y} & 1-y\bar{y}%
\end{array}
\right) .%
\end{equation}
Using these, the Witten ansatz can be rewritten through the singular gauge
transformation
\begin{equation}
g=\left(
\begin{array}
[c]{cc}%
\cos\displaystyle{\frac{\theta}{2}} & e^{-i\varphi}\sin\displaystyle{\frac
{\theta}{2}}\\
-e^{i\varphi}\sin\displaystyle{\frac{\theta}{2}} & \cos\displaystyle{\frac
{\theta}{2}}%
\end{array}
\right)  ={\frac{1}{\sqrt{1+y\bar{y}}}}\left(
\begin{array}
[c]{cc}%
1 & y\\
-\bar{y} & 1
\end{array}
\right)  ,
\end{equation}
which leads to nothing but the gauge configuration of Manton \& Sakai\cite{Manton:2010wu}
\begin{equation}
{A}_{y}^{\mathrm{MS}}={\frac{i}{1+y\bar{y}}}\left(
\begin{array}
[c]{cc}%
-\bar{y}/2 & iH^{\dagger}\\
0 & \bar{y}/2
\end{array}
\right)  ={\frac{1}{1+y\bar{y}}}\left(  -\Phi-i\Lambda\bar{y}\right)
,\quad\left(  \Lambda\equiv{\frac{1}{2}}\sigma_{3},\ \Phi\equiv H^{\dagger
}\sigma_{+}\right)
\end{equation}%
\begin{equation}
{A}_{\bar{y}}^{\mathrm{MS}}={\frac{i}{1+y\bar{y}}}\left(
\begin{array}
[c]{cc}%
y/2 & 0\\
-iH & -y/2
\end{array}
\right)  ={\frac{1}{1+y\bar{y}}}\left(  \bar{\Phi}+i\Lambda y\right)  .
\end{equation}
A generator for rotation around the third axis is
\begin{equation}
x_{2}{\frac{\partial}{\partial x_{1}}}-x_{1}{\frac{\partial}{\partial x_{2}}%
}=i\left(  -y{\frac{\partial}{\partial y}}+\bar{y}{\frac{\partial}%
{\partial\bar{y}}}\right)  ,
\end{equation}
and a generator for the gauge transformation around $\sigma_{3}$ axis is
$i\Lambda$. Using these, the symmetry
equation\cite{Forgacs:1979zs} reads
\begin{align}
\left(  -y{\frac{\partial}{\partial y}}+\bar{y}{\frac{\partial}{\partial
\bar{y}}}\right)  A_{y}+\left[  {\frac{i}{2}}\sigma_{3},A_{y}\right]   &
=A_{y},\\
\left(  -y{\frac{\partial}{\partial y}}+\bar{y}{\frac{\partial}{\partial
\bar{y}}}\right)  A_{\bar{y}}+\left[  {\frac{i}{2}}\sigma_{3},A_{\bar{y}%
}\right]   &  =-A_{\bar{y}}.\nonumber
\end{align}
$A_{y}^{\mathrm{MS}},A_{\bar{y}}^{\mathrm{MS}}$ satisfies the above equation
and thus are the equivariant configurations.

Assuming that these configurations describe the ground state in $S^{2}$ it can
be shown that the pure YM theory in $R^{2}\times S^{2}$ is equivalent with
the YMH theory in $R^{2}$, i.e. SD equation for this model is
\begin{equation}
\left\{
\begin{array}
[c]{l}%
\displaystyle{\frac{8}{(1+y\bar{y})^{2}}}{F}_{z\bar{z}}={F}_{y\bar{y}},\\
{F}_{z\bar{y}}=0,\\
{F}_{\bar{z}y}=0,
\end{array}
\right.
\end{equation}
and substituting the above configuration, each turns into the respective BPS
equation
\begin{equation}
\left\{
\begin{array}
[c]{l}%
F_{z\bar{z}}=\displaystyle{\frac{1}{8}}(2i\Lambda-[\Phi,\bar{\Phi}]),\\
D_{z}\bar{\Phi}=0,\\
D_{\bar{z}}\Phi=0.
\end{array}
\right.
\end{equation}
Here $z,\bar{z}$ are the coordinates on $R^{2}$, and $F_{z\bar{z}}%
=\partial_{z}A_{\bar{z}}-\partial_{\bar{z}}A_{z}+[A_{z},A_{\bar{z}}]$,
$D_{z}\bar{\Phi}=\partial_{z}\bar{\Phi}+[A_{z},\bar{\Phi}]$, $D_{\bar{z}}%
\Phi=\partial_{\bar{z}}\Phi+[A_{\bar{z}},\Phi]$. Namely, gauge field on
$S^{2}$ is recognized as Higgs field on $R^{2}$. On the other
hand, $A_{z},A_{\bar{z}}$ are the gauge fields on the real space
$R^{2}$, they are not transformed by $\Lambda$, i.e.
\begin{equation}
\lbrack\Lambda,A_{z}]=[\Lambda,A_{\bar{z}}]=0.
\end{equation}
Then,
\begin{equation}
A_{z}=\left(
\begin{array}
[c]{cc}%
A_{z}^{L} & 0\\
0 & A_{z}^{R}%
\end{array}
\right)  ,\quad A_{\bar{z}}=\left(
\begin{array}
[c]{cc}%
A_{\bar{z}}^{L} & 0\\
0 & A_{\bar{z}}^{R}%
\end{array}
\right)  ,
\end{equation}
and SU(2$N$) gauge symmetry reduces to $\mathrm{S}\left(  \mathrm{U}%
(N)_{L}\times\mathrm{U}(N)_{R}\right)  $. As a consequence, the BPS equation
is reduced further to
\begin{equation}
\left\{
\begin{array}
[c]{l}%
F_{z\bar{z}}^{L}=\displaystyle{\frac{1}{8}}(-1+H^{\dagger}H),\quad F_{z\bar
{z}}^{R}=\displaystyle{\frac{1}{8}}(1-HH^{\dagger}),\\
D_{z}H^{\dagger}=0,\quad D_{\bar{z}}H=0.
\end{array}
\right.
\end{equation}

\section{Equivariance on $R^{2}\times Z_{2}$ Model}

In this section we consider the case where the extra dimensional space is
noncommutative $Z_{2}$. As a gauge symmetry we consider SU($N$). In order to
make the $Z_{2}$ space invisible, we choose $i\tau_{3}$ as an
equivariant quantity $Q$, of which each component is $N\times N$ SU($N$)
matrix. This matrix describes the $Z_{2}$ space and is
itself the block matrix each block expressing the SU($N$).

The coordinates of noncommutative $Z_{2}$ space are described as\cite{Otsu}
\begin{equation}
w=\left(
\begin{array}
[c]{cc}%
0 & 1\\
0 & 0
\end{array}
\right)  ={\frac{\tau_{1}+i\tau_{2}}{2}},\quad\bar{w}=\left(
\begin{array}
[c]{cc}%
0 & 0\\
1 & 0
\end{array}
\right)  ={\frac{\tau_{1}-i\tau_{2}}{2}}.
\end{equation}
Coordinate transformation for $Z_{2}$ space is discrete, and can
be realized by $i\tau_{1}$, i.e.
\begin{equation}
(i\tau_{1})w(-i\tau_{1})=\left(
\begin{array}
[c]{cc}%
0 & 1\\
1 & 0
\end{array}
\right)  \left(
\begin{array}
[c]{cc}%
0 & 1\\
0 & 0
\end{array}
\right)  \left(
\begin{array}
[c]{cc}%
0 & 1\\
1 & 0
\end{array}
\right)  =\left(
\begin{array}
[c]{cc}%
0 & 0\\
1 & 0
\end{array}
\right)  =\bar{w}.
\end{equation}
If we transform $Q(=i\tau_{3})$ by the $i\tau_1$, we obtain
\begin{equation}
(i\tau_{1})i\tau_{3}(-i\tau_{1})=i\left(
\begin{array}
[c]{cc}%
0 & 1\\
1 & 0
\end{array}
\right)  \left(
\begin{array}
[c]{cc}%
1 & 0\\
0 & -1
\end{array}
\right)  \left(
\begin{array}
[c]{cc}%
0 & 1\\
1 & 0
\end{array}
\right)  =i\left(
\begin{array}
[c]{cc}%
-1 & 0\\
0 & 1
\end{array}
\right)  =-i\tau_{3},
\end{equation}
which changes the sign. 
%
%
Then, we should choose the block components of $i\tau_3$ which are $N\times N$ matrices, so that the sign of each 
component can be changed by a gauge transformation, and $Q$ becomes equivariant.  
For example, when $N=2$, we choose
\begin{align}
Q=\left(
\begin{array}{c|c}
i\sigma_3 & 0\\
\hline
0 & -i\sigma_3
\end{array}\right)=
\left(
\begin{array}{cc|cc}
i & 0 & 0 & 0\\
0 & -i & 0 & 0\\
\hline
0 & 0 & -i & 0\\
0 & 0 & 0 & i
\end{array}\right).
\end{align}
So, using the gauge transformation in terms of
\begin{align}
g=\left(
\begin{array}{c|c}
i\sigma_1 & 0 \\
\hline
0 & i\sigma_1
\end{array}\right),
\end{align}
$Q$ transformed by the spacial rotation ($i\tau_1$) returns to the original one. That is,
\begin{align}
g^\dagger(i\tau_1)Q(-i\tau_1)g&=-g^\dagger Qg\\
&=-\left(
\begin{array}{c|c}
(-i\sigma_1)(i\sigma_3)(i\sigma_1) & 0 \\
\hline
0 & (-i\sigma_1)(-i\sigma_3)(i\sigma_1)
\end{array}\right)  \nonumber \\
&=
-\left(
\begin{array}{c|c}
-i\sigma_3 & 0 \\
\hline
0 & i\sigma_3
\end{array}\right)
=Q. \nonumber
\end{align}
Using this $Q$, we can construct the equivariant gauge field configuration.

As a consequence, 
both the $Z_2$ space and this discrete part of the gauge
transformation become invisible, if the ground state is described by the $Q$.

We define the differential operators by the graded
commutators\cite{Coquereaux}
\begin{align}
\partial_{w}f  &  =[\bar{w},f\}=\bar{w}f-(-1)^{[f]}f\bar{w},\\
\partial_{\bar{w}}f  &  =[w,f\}=wf-(-1)^{[f]}fw.\nonumber
\end{align}
(about the definition, see the Appendix), where $[f]$ is $+1,$ when $f$ is
even, $-1$ when $f$ is odd matrix.

As $Q^{2}=-1$, vectors that can be constructed from $Q$ are $\partial
_{w(\bar{w})}Q$ and $Q\partial_{w(\bar{w})}Q$. As $Q$ is an even matrix the
differential operator is a commutator, expressing in terms of $\tau_{i}$ we
have
\begin{equation}
\partial_{w}Q={\frac{1}{2}}[\tau_{1}-i\tau_{2},i\tau_{3}]=\left(
\begin{array}
[c]{cc}%
0 & 0\\
2i & 0
\end{array}
\right)  ,
\end{equation}%
\begin{equation}
Q\partial_{w}Q=\tau_{3}(\tau_{1}+i\tau_{2})=i\tau_{2}+\tau_{1}=\left(
\begin{array}
[c]{cc}%
0 & 0\\
2 & 0
\end{array}
\right)  ,
\end{equation}
similarly
\begin{equation}
\partial_{\bar{w}}Q={\frac{1}{2}}[\tau_{1}+i\tau_{2},i\tau_{3}]=\left(
\begin{array}
[c]{cc}%
0 & -2i\\
0 & 0
\end{array}
\right)  ,
\end{equation}%
\begin{equation}
Q\partial_{\bar{w}}Q=-\tau_{3}(\tau_{1}-i\tau_{2})=-i\tau_{2}+\tau_{1}=\left(
\begin{array}
[c]{cc}%
0 & 2\\
0 & 0
\end{array}
\right)  .
\end{equation}
Thus, as the gauge configuration in $Z_{2}$ space, we have the 
odd matrix
\begin{equation}
A_{w}=\left(
\begin{array}
[c]{cc}%
0 & 0\\
H & 0
\end{array}
\right)  ,\ A_{\bar{w}}=\left(
\begin{array}
[c]{cc}%
0 & H^{\dagger}\\
0 & 0
\end{array}
\right)  ,
\end{equation}
which are equivariant under gauge, coordinate and spin
transformations like in Eq.(1). In this case, the spin transformation is given as
\begin{equation}
A_{w(\bar w)}\rightarrow -A_{\bar w(w)},
\end{equation}
in response to the coordinate transformation, because $A_{w(\bar w)}$ are the vector
fields.

Next we define the field strength in $Z_{2}$ space. The field strength is
usually defined as $F_{\mu\nu}=[D_{\mu},D_{\nu}]$. We extend this in $Z_{2}$
space as
\begin{equation}
F_{XY}=[D_{X},D_{Y}\}\equiv D_{X}D_{Y}-(-1)^{[X][Y]}D_{Y}D_{X} .%
\qquad(X,Y=z,\bar{z},w,\bar{w})
\end{equation}

From the definition of graded commutator and Jacobi identity
\begin{equation}
\lbrack f,g\}=fg-(-1)^{[f][g]}gf,
\end{equation}%
\begin{equation}
(-1)^{[A][C]}[A,[B,C\}\}+(-1)^{[A][B]}[B,[C,A\}\}+(-1)^{[C][B]}[C,[A,B\}\}=0,
\end{equation}
we have
\begin{align}
D_{w}D_{\bar{w}}f  &  =(\partial_{w}+A_{w})(\partial_{\bar{w}}f+A_{\bar{w}%
}f)+A_{w}(\partial_{\bar{w}}f)+A_{w}A_{\bar{w}}f\\
&  =\partial_{w}(\partial_{\bar{w}}f)+(\partial_{w}A_{\bar{w}})f-A_{\bar{w}%
}(\partial_{w}f)+A_{w}(\partial_{\bar{w}}f)+A_{w}A_{\bar{w}}f,%
\nonumber
\end{align}%
\begin{equation}
D_{\bar{w}}D_{w}f=\partial_{\bar{w}}(\partial_{w}f)+(\partial_{\bar{w}}%
A_{w})f-A_{w}(\partial_{\bar{w}}f)+A_{\bar{w}}(\partial_{w}f)+A_{\bar{w}}%
A_{w}f.
\end{equation}
Taking into account $\partial_{\bar{w}}\partial_{w}f+\partial_{w}%
\partial_{\bar{w}}f=0$, we have
\begin{align}
D_{w}D_{\bar{w}}f+D_{\bar{w}}D_{w}f  &  =(\partial_{w}A_{\bar{w}}%
)f+(\partial_{\bar{w}}A_{w})f+A_{w}A_{\bar{w}}f+A_{\bar{w}}A_{w}f\\
&  =(\partial_{w}A_{\bar{w}}+\partial_{\bar{w}}A_{w}+\{A_{w},A_{\bar{w}%
}\})f.\nonumber
\end{align}
As a result, $F_{w\bar{w}}$ can be written as $\{D_{w},D_{\bar{w}}\}$.

Next, we consider $D_{w}D_{\bar{z}}f$ and $D_{\bar{z}}D_{w}f$. Taking into
account that $A_{\bar{z}}$ is an even matrix, we calculate
\begin{align}
D_{w}D_{\bar{z}}f  &  =(\partial_{w}+A_{w})(\partial_{\bar{z}}f+A_{\bar{z}%
}f)\\
&  =\partial_{w}(\partial_{\bar{z}}f)+(\partial_{w}A_{\bar{z}})f+A_{\bar{z}%
}\partial_{w}f+A_{w}\partial_{\bar{z}}f+A_{w}A_{\bar{z}}f,%
\nonumber
\end{align}%
\begin{align}
D_{\bar{z}}D_{w}f  &  =(\partial_{\bar{z}}+A_{\bar{z}})(\partial_{w}%
f+A_{w}f)\nonumber\\
&  =\partial_{\bar{z}}(\partial_{w}f)+(\partial_{\bar{z}}A_{w})f+A_{w}%
\partial_{\bar{z}}f+A_{\bar{z}}\partial_{w}f+A_{\bar{z}}A_{w}f,%
\end{align}
and we can consider of $[D_{w},D_{\bar{z}}]$ as a field strength, i.e.
\begin{align}
D_{w}D_{\bar{z}}f-D_{\bar{z}}D_{w}f  &  =(\partial_{w}A_{\bar{z}}%
)f-(\partial_{\bar{z}}A_{w})f+A_{w}A_{\bar{z}}f-A_{\bar{z}}A_{w}f\\
&  =(\partial_{w}A_{\bar{z}}+\partial_{\bar{z}}A_{w}+[A_{w},A_{\bar{z}%
}])f ,\nonumber
\end{align}
thus $[D_{w},D_{\bar{z}}]=F_{w\bar{z}}$. As for $F_{z\bar{z}}$, it is the
ordinary field strength on $R^{2}$, thus $F_{z\bar{z}}=[D_{z},D_{\bar{z}}]$.

From these arguments, it is appropriate to define
\begin{equation}
F_{XY}=[D_{X},D_{Y}\}\equiv D_{X}D_{Y}-(-1)^{[X][Y]}D_{Y}D_{X}.\qquad
(X,Y=z,\bar{z},w,\bar{w})
\end{equation}

Now we calculate the field strength for the gauge configurations 
\begin{equation}
A_{w}=\left(
\begin{array}
[c]{cc}%
0 & 0\\
H & 0
\end{array}
\right)  ,\ A_{\bar{w}}=\left(
\begin{array}
[c]{cc}%
0 & H^{\dagger}\\
0 & 0
\end{array}
\right)  .
\end{equation}
For example
\begin{align}
F_{\bar{z}w}  &  =\partial_{\bar{z}}A_{w}-\partial_{w}A_{\bar{z}}+\left[
A_{\bar{z}},A_{w}\right] \\
&  =\partial_{\bar{z}}\left(
\begin{array}
[c]{cc}%
0 & 0\\
H & 0
\end{array}
\right)  -\partial_{w}\left(
\begin{array}
[c]{cc}%
A_{\bar{z}}^{L} & 0\\
0 & A_{\bar{z}}^{R}%
\end{array}
\right)  +\left[  \left(
\begin{array}
[c]{cc}%
A_{\bar{z}}^{L} & 0\\
0 & A_{\bar{z}}^{R}%
\end{array}
\right)  ,\left(
\begin{array}
[c]{cc}%
0 & 0\\
H & 0
\end{array}
\right)  \right] \nonumber\\
&  =\left(
\begin{array}
[c]{cc}%
0 & 0\\
\partial_{\bar{z}}\phi-\phi A_{\bar{z}}^{L}+A_{\bar{z}}^{R}\phi & 0
\end{array}
\right)  \qquad(\phi=H+1)\nonumber\\
&  =\left(
\begin{array}
[c]{cc}%
0 & 0\\
D_{\bar{z}}\phi & 0
\end{array}
\right)  ,\nonumber
\end{align}
and
\begin{align}
F_{w\bar{w}}=  &  \partial_{w}A_{\bar{w}}+\partial_{\bar{w}}A_{w}+\left\{
A_{w},A_{\bar{w}}\right\} \\
=  &  \left\{  \left(
\begin{array}
[c]{cc}%
0 & 0\\
1 & 0
\end{array}
\right)  ,\left(
\begin{array}
[c]{cc}%
0 & H^{\dagger}\\
0 & 0
\end{array}
\right)  \right\}  +\left\{  \left(
\begin{array}
[c]{cc}%
0 & 1\\
0 & 0
\end{array}
\right)  ,\left(
\begin{array}
[c]{cc}%
0 & 0\\
H & 0
\end{array}
\right)  \right\} \nonumber\\
&  +\left\{  \left(
\begin{array}
[c]{cc}%
0 & 0\\
H & 0
\end{array}
\right)  ,\left(
\begin{array}
[c]{cc}%
0 & H^{\dagger}\\
0 & 0
\end{array}
\right)  \right\} \nonumber\\
=  &  \left(
\begin{array}
[c]{cc}%
\phi^{\dagger}\phi-1 & 0\\
0 & \phi\phi^{\dagger}-1
\end{array}
\right)  .\nonumber
\end{align}

Next, in order to consider the (anti-)self dual equation we define
$\widetilde{F}_{XY}$. On $R^{2}\times Z_{2}$, since definition of $F_{XY}$ is
different from the usual one, in particular $F_{\bar{w}w}$ is defined as an
anticommutator, we cannot write
\begin{equation}
\widetilde{F}_{\mu\nu}={\frac{1}{2}}\epsilon_{\mu\nu\lambda\rho}F^{\lambda
\rho},
\end{equation}
as in the case of $R^{4}$. Although $F_{\bar{w}w}$ was defined as
\begin{align}
F_{\bar{w}w}  &  =\partial_{\bar{w}}A_{w}+\partial_{w}A_{\bar{w}}+\{A_{\bar
{w}},A_{w}\}\\
&  =\left\{  \left(
\begin{array}
[c]{cc}%
0 & 1\\
0 & 0
\end{array}
\right)  ,A_{w}\right\}  +\left\{  \left(
\begin{array}
[c]{cc}%
0 & 0\\
1 & 0
\end{array}
\right)  ,A_{\bar{w}}\right\}  +\{A_{\bar{w}},A_{w}\},\nonumber
\end{align}
we can  redefine this as a commutator using $\tau_{3}$, as follows
\begin{align}
F_{\bar{w}w}  &  =\partial_{\bar{w}}A_{w}+\partial_{w}A_{\bar{w}}+\{A_{\bar
{w}},A_{w}\}\\
&  =\left[  \left(
\begin{array}
[c]{cc}%
0 & 1\\
0 & 0
\end{array}
\right)  ,A_{w}\right]  \tau_{3}-\left[  \left(
\begin{array}
[c]{cc}%
0 & 0\\
1 & 0
\end{array}
\right)  ,A_{\bar{w}}\right]  \tau_{3}+[A_{\bar{w}},A_{w}]\tau_{3}.
\end{align}
Consequently, if we multiply this equation by $\tau_{3}$ we
obtain the usual expression for field strength. Taking into account that
$A_{z},A_{\bar{z}}$ are dual to $A_{w},A_{\bar{w}}$, $\tilde{F}$ can be
defined by taking dual and multiplying by $\tau_{3}$ (for details, see the
Appendix). Thus the (anti) self dual equations can be written as
\begin{equation}
F_{\bar{z}w}=0,\quad F_{w\bar{w}}=\pm F_{z\bar{z}}\tau_{3},\quad\mathrm{etc.}%
\end{equation}
Substituting these equations into the ansatz, each BPS equations are
equivalent to
\begin{equation}
D_{\bar{z}}\phi=0,\quad F_{z\bar{z}}=\phi\phi^{\dagger}-1,\quad\mathrm{etc.}%
\end{equation}%
\begin{equation}
\left\{
\begin{array}
[c]{l}%
F_{z\bar{z}}^{L}=\phi^{\dagger}\phi-1,\quad F_{z\bar{z}}^{R}=-\phi
\phi^{\dagger}+1\\
D_{z}\phi^{\dagger}=0,\quad D_{\bar{z}}\phi=0
\end{array}
\right.
\end{equation}
which are consistent with \cite{Otsu}. \newline\newline

\section{Discussion}

We have been discussing the YM theory on $R^{2}\times Z_{2}$. In \cite{Otsu},
by use of differential form, we have seen that the self dual equation for YM
theory on $R^{2}\times Z_{2}$ is equivalent to the BPS equation for YMH model
on $R^{2}$. In the present paper, by explicitly introducing the coordinates on
$Z_{2}$ space, we show the equivalence of YM theory on $R^{2}\times Z_{2}$ and
YMH model on $R^{2}$, based on the idea of equivariance. This argument is the
same as in the case of showing the equivalence of YM theory on $R^{2}\times
S^{2}$ and the YMH model on $R^{2}$.

Moreover, because the $Z_{2}$ space is discrete, the required equivariance is
among the discrete part of gauge symmetry and it is unnecessary to consider
the larger gauge symmetry. In other words, we could remain with the same gauge 
symmetry and this is probably the simplest model of this kind.

The equivariant gauge configuration is
described by the basic equivariant quantity $Q$ exactly as in the case of $S^{2}$ space. As we have
succeeded in introducing $Q$ in $Z_{2}$ space, we were able to construct gauge
configuration for the ground state out of $Q$. In other words, we succeeded in
introducing the Witten ansatz in the noncommutative $Z_{2}$ space.

As stated before, we have seen in \cite{Otsu} that YM theory on $R^{2}\times
Z_{2}$ space is equivalent to YMH model on $R^{2}$ space. The
argument was based on the differential forms, and we were able to construct
the differential forms on $Z_{2}$ space using the matrices. On the other hand,
in this paper, we have explicitly introduced the coordinates in $Z_{2}$ space,
and consistently derived the same conclusion as in \cite{Otsu}. The connection
of the theory based on differential forms and present theory based on the
explicit coordinate is not clear, because we have not been able to construct
differential forms for $Z_{2}$ space. We will discuss this point in the future publication.

\appendix
\section*{Appendix}
\section{Differential Operators on $Z_{2}$ Space}

Let $x_{a}$ be the coordinates in $R^{2}$, and $y_{\alpha}$ the coordinates of
the curved space. The metric in the curved space can be expressed as
$g_{\alpha\beta}=e_{\alpha}^{a}(y)e_{\beta}^{a}(y)$, and
\begin{equation}
{\frac{\partial}{\partial y_{\alpha}}}=e_{\alpha}^{a}(y){\frac{\partial
}{\partial x_{a}}}.
\end{equation}
Poisson bracket is defined as
\begin{equation}
\{f,g\}_{P}\equiv\theta^{\alpha\beta}(y){\frac{\partial f}{\partial y^{\alpha
}}}{\ \frac{\partial g}{\partial y^{\beta}}},
\end{equation}
where
\begin{equation}
\theta^{\alpha\beta}(y)=\theta(y)\epsilon^{\alpha\beta},\quad\theta
(x)={\frac{1}{\sqrt{g(y)}}}.
\end{equation}
From these definitions, we find
\begin{equation}
\{y^{\alpha},y^{\beta}\}_P=\theta^{\alpha\beta},\quad{\frac{\partial}{\partial
y^{\alpha}}}=\theta_{\alpha\beta}^{-1}\{y^{\beta},\ \}_{P}.
\end{equation}

If we consider the ``correspondence principle'' for the curved space,
 Poisson bracket can be replaced by commutator\cite{Pinzul:2005nh}
\begin{equation}
\lbrack w^{\mu},w^{\nu}]=\theta_{F}^{\mu\nu},
\end{equation}
where $w_\mu$'s are coordinates of the fuzzified space. 
Then, differential operators can be written as $(\theta_{F}^{\mu\nu})^{-1}
[w^{\nu},\ ]$. 
Now, regarding $Z_{2}$ space as the fuzzified space of a certain curved space,
we have to find $\theta_{F}^{\mu\nu}$ in our case. 

In our case,
\begin{equation}
\lbrack w,\bar{w}]=2i[w_{1},w_{2}]=\tau_{3},
\end{equation}
then, we find
\begin{equation}
\theta_{F}^{\mu\nu}=\tau_{3}\epsilon^{\mu\nu}.
\end{equation}
Therefore, the differential operators can be written as
\begin{equation}
\partial_{w}=-\tau_{3}[\bar{w},\ ],\ \partial_{\bar{w}}=\tau_{3}[w,\ ],
\end{equation}
and the self dual equation is
\begin{equation}
F_{\bar{z}z}=\tau_{3}F_{\bar{w}w}.
\end{equation}
This fact is interpreted as follows: in $Z_2$ space, the factor $\tau_3$ is necessary for the definition of differential operators,
so that the Jacobi identity and the Leibniz rule are satisfied. This suggests that the $Z_2$ space is a certain curved space, and so 
this space possesses a measure different from that of the flat space. Therefore, $\tau_3$ should appear in the self dual equation 
as a correction factor of the measure and volume element. This fact has been confirmed 
in the previous paper by the method using the
differential fom that does not depend on the coordinates of $Z_2$ space.

On the other hand, these differential operators can be rewritten in terms of
the graded commutators. Let $A$ be any $2\times2$ matrix. $A$ can be decomposed
as
\begin{equation}
A=A_{e}+A_{o},
\end{equation}
where $A_{e}$ is even matrix, and $A_{o}$ is odd one. Namely,
\begin{equation}
A_{e}=\left(
\begin{array}
[c]{cc}%
a & 0\\
0 & b
\end{array}
\right)  ,\quad A_{o}=\left(
\begin{array}
[c]{cc}%
0 & c\\
d & 0
\end{array}
\right)  .
\end{equation}
Then, it is easily found that the differentials of $A$ are
\begin{align}
\partial_{w}A  &  =-\tau_{3}[\bar{w},A]=-\tau_{3}[\bar{w},A_{e}]-\tau_{3}%
[\bar{w},A_{o}]\\
&  =\left(
\begin{array}
[c]{cc}%
0 & 0\\
a-b & 0
\end{array}
\right)  +\left(
\begin{array}
[c]{cc}%
c & 0\\
0 & c
\end{array}
\right)  ,
\end{align}%
\begin{align}
\partial_{\bar{w}}A  &  =\tau_{3}[w,A]=\tau_{3}[w,A_{e}]+\tau_{3}[w,A_{o}]\\
&  =\left(
\begin{array}
[c]{cc}%
0 & b-a\\
0 & 0
\end{array}
\right)  +\left(
\begin{array}
[c]{cc}%
d & 0\\
0 & d
\end{array}
\right)  .\nonumber
\end{align}
These relations can also be realized in terms of the graded commutators as
\begin{align}
\partial_{w}A  &  =[\bar{w},A\}=\bar{w}A-(-1)^{[A]}A\bar{w},\\
\partial_{\bar{w}}A  &  =[w,A\}=wA-(-1)^{[A]}Aw.\nonumber
\end{align}
We adopt this definition in this article.

\section*{Acknowledgments}

We would like to thank Akihiro Nakayama for his support and hospitality.

\end{document}